\documentclass[sigconf, nonacm]{acmart}

\newcommand\vldbdoi{XX.XX/XXX.XX}

\newcommand\vldbavailabilityurl{http://vldb.org/pvldb/format_vol14.html}
\newcommand\vldbpagestyle{plain} 

\begin{document}
\title{ViralFP: A webserver of viral fusion proteins}

%%
%% The "author" command and its associated commands are used to define the authors and their affiliations.
\author{Pedro Moreira}
\affiliation{%
  \institution{Escola de Engenharia da Universidade do Minho}
  \streetaddress{Rua da Universidade}
  \city{Braga}
  \state{Portugal}
  \postcode{4710-057}
}
\email{pg38277@alunos.uminho.pt*}

\author{Ana Marta Sequeira}
\affiliation{%
  \institution{Escola de Engenharia da Universidade do Minho}
  \streetaddress{Rua da Universidade}
  \city{Braga}
  \state{Portugal}
  \postcode{4710-057}
}

\author{Sara Pereira}
\affiliation{%
  \institution{Escola de Engenharia da Universidade do Minho}
  \streetaddress{Rua da Universidade}
  \city{Braga}
  \state{Portugal}
  \postcode{4710-057}
}

\author{Rúben Rodrigues}
\affiliation{%
  \institution{Escola de Engenharia da Universidade do Minho}
  \streetaddress{Rua da Universidade}
  \city{Braga}
  \state{Portugal}
  \postcode{4710-057}
}

%\orcid{0000-0002-1825-0097}
\author{Miguel Rocha}
\affiliation{%
  \institution{Escola de Engenharia da Universidade do Minho}
  \streetaddress{Rua da Universidade}
  \city{Braga}
  \state{Portugal}
  \postcode{4710-057}
}
\email{mrocha@di.uminho.pt}

%\orcid{0000-0001-5109-3700}
\author{Diana Lousa}
\affiliation{%
  \institution{ITQB NOVA, Universidade Nova de Lisboa}
  \streetaddress{Av. da República}
  \city{Oeiras}
  \state{Portugal}
  \postcode{2780-157}
}
\email{dlousa@itqb.unl.pt}

\begin{abstract}
Viral fusion proteins are attached to the membrane of enveloped viruses (a group that includes Coronaviruses, Dengue, HIV and Influenza) and catalyze fusion between the viral and host membranes, enabling the virus to insert its genetic material into the host cell. Given the importance of these biomolecules, this work presents a centralized database containing the most relevant information on viral fusion proteins, available through a free-to-use web server accessible through the URL \url{https://viralfp.bio.di.uminho.pt/}. This web application contains several bioinformatic tools, such as Clustal sequence alignment and Weblogo, including as well a machine learning-based tool capable of predicting the location of fusion peptides (the component of fusion proteins that inserts into the host's cell membrane) within the fusion protein sequence. Given the crucial role of these proteins in viral infection, their importance as natural targets of our immune system and their potential as therapeutic targets, this web application aims to foster our ability to fight pathogenic viruses.   \par
\vspace{.3cm}
\end{abstract}

\maketitle

\pagestyle{\vldbpagestyle}
%\begingroup\small\noindent\raggedright\textbf{PVLDB Reference Format:}\\
%\vldbauthors. \vldbtitle. PVLDB, \vldbvolume(\vldbissue): \vldbpages, \vldbyear.\\
%\href{https://doi.org/\vldbdoi}{doi:\vldbdoi}
%\endgroup
\begingroup
%\renewcommand\thefootnote{}\footnote{\noindent
%This work is licensed under the Creative Commons BY-NC-ND 4.0 International License. Visit \url{https://creativecommons.org/licenses/by-nc-nd/4.0/} to view a copy of this license. For any use beyond those covered by this license, obtain permission by emailing \href{mailto:info@vldb.org}{info@vldb.org}. Copyright is held by the owner/author(s). Publication rights licensed to the VLDB Endowment. \\
%\raggedright Proceedings of the VLDB Endowment, Vol. \vldbvolume, No. \vldbissue\ %
%ISSN 2150-8097. \\
%\href{https://doi.org/\vldbdoi}{doi:\vldbdoi} \\
%}%\addtocounter{footnote}{-1}\endgroup

\ifdefempty{\vldbavailabilityurl}{}{
%\vspace{.3cm}
%\begingroup\small\noindent\raggedright\textbf{PVLDB Artifact Availability:}\\
%The source code, data, and/or other artifacts have been made available at \url{\vldbavailabilityurl}.
%\endgroup
}


\section{Introduction}

Enveloped viruses, a group of viruses that includes SARS-CoV-2, Dengue, Influenza and HIV, are characterized by having an exterior lipid envelope that contains at least one type of glycoprotein on its surface \citep[Preface I]{Marsh2005} \citep{Marsh2006} \citep{Weissenhorn2007}. These proteins play an essential role during the virus entry into the host cell, enabling the binding of the virus to receptors on the host cell and catalysing the fusion between the viral envelope and the host cell membrane. 

In some viruses, a single protein can induce the binding to the host receptor and catalyze the fusion reaction, whereas other viruses require multiple proteins for these tasks. Viral fusion proteins (VFP) are key players in a process called membrane fusion, through which the virus inserts its genetic material into the host cell, enabling it to replicate \citep{Apellaniz2014} \citep{Harrison2015}.

During membrane fusion, the viral envelope and the host cellular membrane are brought into close contact in order to merge into a single membrane \citep{Baquero2013} \citep{White2008}. According to the most widely accepted membrane fusion model, this process is initiated when the VFP binds to the host receptor. Triggered by an external stimulus, the protein becomes extended and inserts its fusion peptide (FPep) into the host membrane. The VFP folds back, which leads to the approximation of the two membranes. After this step, the mechanism passes through a hemifusion intermediate, which encompasses the fusion of the outer leflets of the membranes without the contact of the distal leaflets. This hemifused step forms a stalk-like structure, diminishing the contact area to a minimum, lowering the work required to overcome hydration force repulsion. Finally, fusion of the unattached inner leaflets occurs, which starts to form a hydrophilic pore that will eventually enlarge, connecting the 2 initially separated intermembrane volumes \citep{Apellaniz2014} \citep{Harrison2015} \citep{Hughson1995} \citep{Teissier2007}.

VFPs can assume multiple possible forms depending on their corresponding virus. Currently, the consensus is that VFPs can be grouped into 3 classes: I, II and III, each with their own characteristics, such as their prefusion, pre-hairpin and post-fusion structures, orientation within virus’ membrane, types of possible triggering mechanisms and location of their FPep \citep{Apellaniz2014} \citep{Epand2003} \citep{Harrison2015} \citep{White2008}.

The FPep is a conserved sequence within VFPs of the same viral family that can insert into the host membrane, allowing the fusion of the viral envelope with the host cell membrane. This peptide can promote the reaction of lipid mixing of membrane vesicles by itself, even if it is detached from the rest of the VFP \citep{Apellaniz2014} \citep{Epand2003}.

The relevance of VFP and their FPep for membrane fusion highlights them as potential therapeutic targets \citep{Apellaniz2014} \citep{White2008}. As an example, the therapeutic application of VFPs as a target can be found in the defining pandemic of 2020: COVID-19. Being an enveloped virus, one of the key therapeutic targets of SARS-CoV-2 is its VFP, the spike (S) protein \citep{li2020targeting}. The most promising and effective vaccines under development or already administrated for this virus are based on the S protein’s mRNA. This group of vaccines include the Moderna’s mRNA-1273 \citep{jackson2020mrna} and the Pfizer-BioNTech’s BNT162b2 mRNA \citep{polack2020safety}. In both, the mRNA molecule is inserted into a lipid nanocarrier that stabilizes the mRNA and delivers it into human cells. The mRNA once inside a cell is translated to produce the S protein in its prefusion conformation. The presence of those proteins in the human cells induces an immune response on the organism, which will produce antibodies that can target those proteins. 

The known information for a given VFP can be found in biological databases, like UniProt \citep{10.1093/nar/gkaa1100}, NCBI Protein \citep{geer2010ncbi} and PDB \citep{Burley2019}. However, a VFP oriented repository that contains VFP related information does not exist, forcing researchers to search several databases when they need to have comprehensive characterization of one or several of these proteins. 

\begin{figure*}
    \centering
    \includegraphics{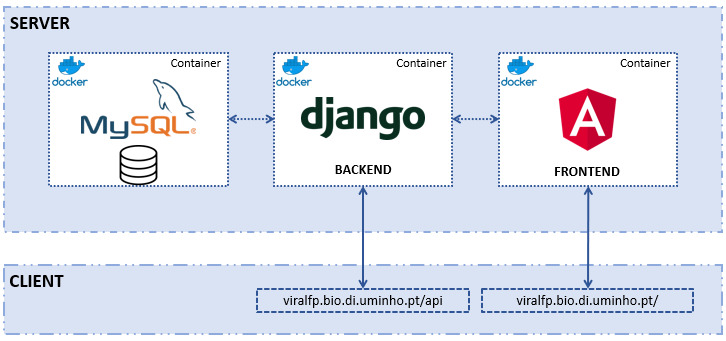}
    \caption{Structure of the app: within the deployment server there are the 3 Docker containers: the MySQL, with the relational database; the Django (back-end) and the Angular (front-end). The latter two can be accessed by a web client (e.g.,  browser or any application that a consumes RESTfull API).} 
    \label{SchemeApp}
\end{figure*}

To better understand how VFPs and their components are characterized, some bioinformatics tools can be used, like sequence alignment (e.g.,  BLAST \citep{johnson2008ncbi}, Clustal \citep{sievers2014clustal} and HMMER \citep{finn2011hmmer}), that allow researchers to discover a protein’s conserved residues, as well as other pieces of information regarding their features. These analyses can provide valuable insights about how VFPs operate. 

Researchers have also been using machine learning (ML) to study VFPs and their inhibitors. Several promising tools have been developed using ML models trained using VFP and their features, allowing the prediction of FPeps location within VFP sequences \citep{Wu2016} \citep{Wu2019} and the prediction of inhibitors for different VFPs to potentially be used in therapies of some diseases \citep{Xu2015}. 

In this context, the main objective of the current work is to develop a CRUD (Create, Read, Update and Delete) web application, named ViralFP, that is capable of storing, retrieving, displaying, and analysing the gathered data on enveloped viruses (through information gathered from NCBI Taxonomy \citep{federhen2012ncbi} entries) and their VFPs, as well as to enable the data navigation through a user-friendly interface. The web application is freely available and aims to be useful for all elements of the scientific community who are working with VFPs and FPeps. The underlying data are stored in a relational database, which contains relevant information regarding over five hundred VFPs. Manual curation of the database information was performed by reviewing the original protein and taxonomy repositories' data entries, as well researching through scientific publications.The back-end component of the application is built using Django and Django REST API library, whilst its front-end is built using Angular technology, a Typescript framework to build web applications. 

This web application also includes a set of selected bioinformatics tools, such as sequence alignment tools (like BLAST and Clustal) and ML models capable to detect FPeps within the VFP sequence.

\section{Materials and Methods}

The structure of the ViralFP application can be seen in the Figure \ref{SchemeApp}. There are three distinct components, each with its own Docker container: 
\begin{itemize}
  \item The MySQL database - stores all the data that the server requires.
  \item Django back-end, with Django REST framework - enables the access and management of the database, besides containing several data views that will be called by the front-end.
  \item Angular front-end - the application designed to be used by the end user that displays the database information and allows the execution of several tools.
\end{itemize}

In the next sections, a deeper explanation for each of those components will be made.

\subsection{Database}

\begin{figure*}
    \centering
    \includegraphics{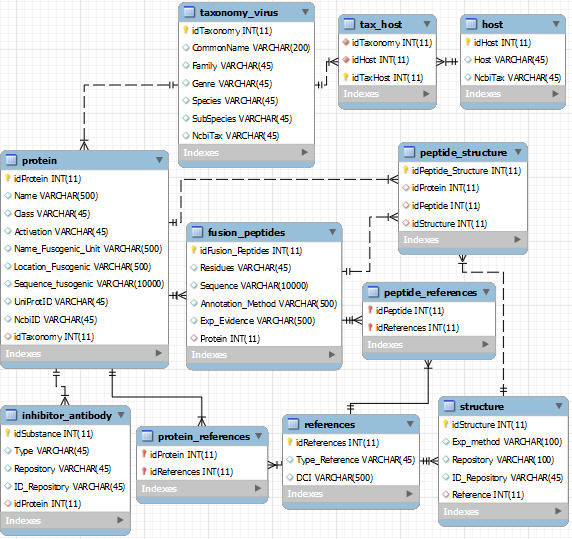}
    \caption{Structure of the relational database, showing its entities and relationships} 
    \label{DB}
\end{figure*}

The VFP / FPep data present in the MySQL database was retrieved from UniProt, NCBI Protein and PDB entries, whilst the taxonomy information was originally from NCBI Taxonomy. Its relational structure, names of the tables and attributes, as well as the type of data stored in each attribute can be seen on the scheme present in figure \ref{DB}.

The core table of this database is called \texttt{Protein} and contains the information regarding each VFP’s name, class, type of activation, fusogenic peptide description (its name, location within the protein and sequence), as well as having the UniProt and NCBI Protein entries for the protein. 

Another important table is the \texttt{Fusion Peptide}, which describes the location of this peptide within the VFP sequence (attribute \texttt{Residues}), sequence and some attributes regarding annotation method and experimental evidence. 

The last key table is the \texttt{Taxonomy}, which has all the relevant taxonomy of the virus to which the protein belongs to (family, genus, species and subspecies / strain), as well as having the virus’ NCBI Taxonomy entry.

Additional tables within the database include a table with the possible structures of the VFP (through PDB entries); a table for the virus' hosts, which has the common name and NCBI Taxonomy of each virus’ host organism; an inhibitor / antibody table, that contains information regarding those types of protein that are associated with a particular VFP, and, finally, tables for additional bibliographical references.

The data present in the database was manually curated by verifying the original repositories of the data entries. Finally, the inhibitors and antibodies entries for each VFP were filled using \textit{BioPython} \citep{Cock2009}for the Pubmed automated searches, and \textit{pypdb} \citep{Gilpin2016} for the PDB searches. 

\subsection{Back-end}

The main purpose of the back-end is to enable the access to the data from the MySQL database. This application component converts data from SQL queries into JSON objects. Those objects can be viewed and managed by a web client, which in this case implements the front-end. The back-end has a RESTful API which enables the submission or modification of data stored in the MySQL database.

One view for each database table was built with the same name of the respective table. Some of those views use serializers that retrieve data from multiple tables related between themselves by foreign keys. Whenever a query request is made, the returned data will be returned as response (also in a JSON format). All these pages will allow to receive requests to insert, update or delete data entries. Besides that, all of those will contain features to allow to filter and / or search data, besides containing pagination methods.

Since bioinformatics packages are commonly written in Python, those tools are integrated within this application layer, that later will be accessed by the front-end by requests. 

Clustal Omega analysis was wrapped in our application using a Clustal console \citep{Sievers2011}. This view will receive the sequences and the additional parameters as part of the request. As result, it performs a Clustal alignment which provides the sequence alignment output as a text response.

To build Weblogos, the packages \textit{weblogo} \footnote{\url{https://weblogo.readthedocs.io/}} and \textit{logomaker} \citep{tareen2020logomaker} are used. 

The ML view allows to perform the FPep prediction for a given sequence using models trained with the in-house \textit{Propythia} package \citep{Sequeira2020}. This package enables the development of ML and deep learning models for classification of peptides. It allows to obtain protein descriptors from their sequences such as physicochemical features, and includes methods for data preprocessing, feature selection, dimensionality reduction, clustering and pipelines to train, evaluate and optimize ML and deep learning models \citep{Sequeira2020}. The models used in this application were trained from a dataset generated by extracting the physicochemical features using the Descriptor module of Propythia from a dataset where the negative instances are composed of protein sequences from transmembrane domains and random sequences extracted from VFP sequences after excluding the FPep. Calculated features went through a wrapper feature selection process using a Support Vector Machine (SVM) model, which had the best performance in this dataset. Finally, ML models were trained with the Propythia's $train\_best\_model$ function from the shallow machine learning class. Those models are stored in Pickle files \footnote{The datasets, as well a small tutorial on how the models were trained, can be seen in \url{https://github.com/pdMM11/Dockers/tree/master/Django_/crmapp/ml_models/DATASETS}}. 

If a certain model is selected, the file containing that model will be loaded and used in an instance of the \textit{Propythia}'s ML class. This ML view also requires the selection of the window size and a size for the gap. After defining those parameters, the $predicted\_window$ function of the Propythia is executed to perform the likelihood prediction of the subpeptide present in a given window to be a FPeps. Those probabilities are returned in a JSON object by the view.

\subsection{Front-end}

The front-end was built using a framework called Nebular \footnote{\url{https://akveo.github.io/nebular/}}, which is an open-source Angular User Interface Library that consists in a pre-built modular, customizable and configurable Angular tool. This framework enables the creation of Angular-based web applications and speeds up the development process of a front-end by providing several useful user-friendly and visually appealing web components, like tables and card formats.

\subsubsection{Tables' Pages}

The information of the database is found on the “Tables \& Data” option of the front-end menu. This menu option contains a list of links for the \texttt{Taxonomy}, \texttt{Fusion Protein} and \texttt{Fusion Peptides} tables. All tables are organised in a Tree Grid format: each data entry will correspond to a table line; upon clicking on one row, its related data, which in this app corresponds to links to other tables connected by foreign keys, will appear below. 

\subsubsection{Sequence Alignment Pages}

The application's tools can be grouped together in two categories: sequence alignment and sequence prediction. This section will cover the first group, while the next section handles the latter.

In the Sequence Alignment tools, a page that enables the submission of sequences to the NCBI BLAST was added. This page encompasses parameters that allow the selection of the type of BLAST and the database which BLAST will run against \citep{johnson2008ncbi}. Similarly, the HMMER page allows the selection of a sequence and the database to perform the alignment in the EMBL-EBI portal, showing the results in their web portal \citep{Potter2018}.  

The Weblogo page contains two possible ways to build sequence logos. The first uses the Weblogo 3 portal \footnote{\url{http://weblogo.threeplusone.com/create.cgi}} \citep{Crooks2004}, in which the page will ask a minimum of three sequences, which must be in a FASTA format. Those sequences are submitted to the back-end to perform a Clustal Omega alignment and from that the web portal builds a weblogo provided in PNG or raw data formats. The second option to generate weblogos consists in the usage of the integrated sequence logo generator of the back-end. The PNG output will appear in the same page, on which the user can select the type of colouring provided by the \textit{logomaker} package, as well as the number of stacks per line; the text output will appear in a pop-up page.

The Clustal page alignment will be run in the back-end, in a command line. The user can select parameters such as the format of output. Upon submitting the request, one pop-up will open in the browser to display the alignment in the required format; a second one can appear if the guide tree data was requested.

\subsubsection{Sequence Prediction Pages}

The web application contains two tools relative to sequence prediction: epitope prediction and FPep prediction.

Regarding the epitope prediction (epitopes are binding sites that antibody molecules can attach to) the site will access to the IEDB portal \citep{Vita2014} to obtain an antibody epitope prediction. The page will send a request with the sequence, selected analysis method and window size to the back-end that will redirect to the IEDB API. The results will be displayed in a Smart Table format. Both Bepipred methods' results will appear into 2 different result tables, the rest of them just require one \citep{Vita2018}.

\begin{figure*}
    \centering
    \includegraphics{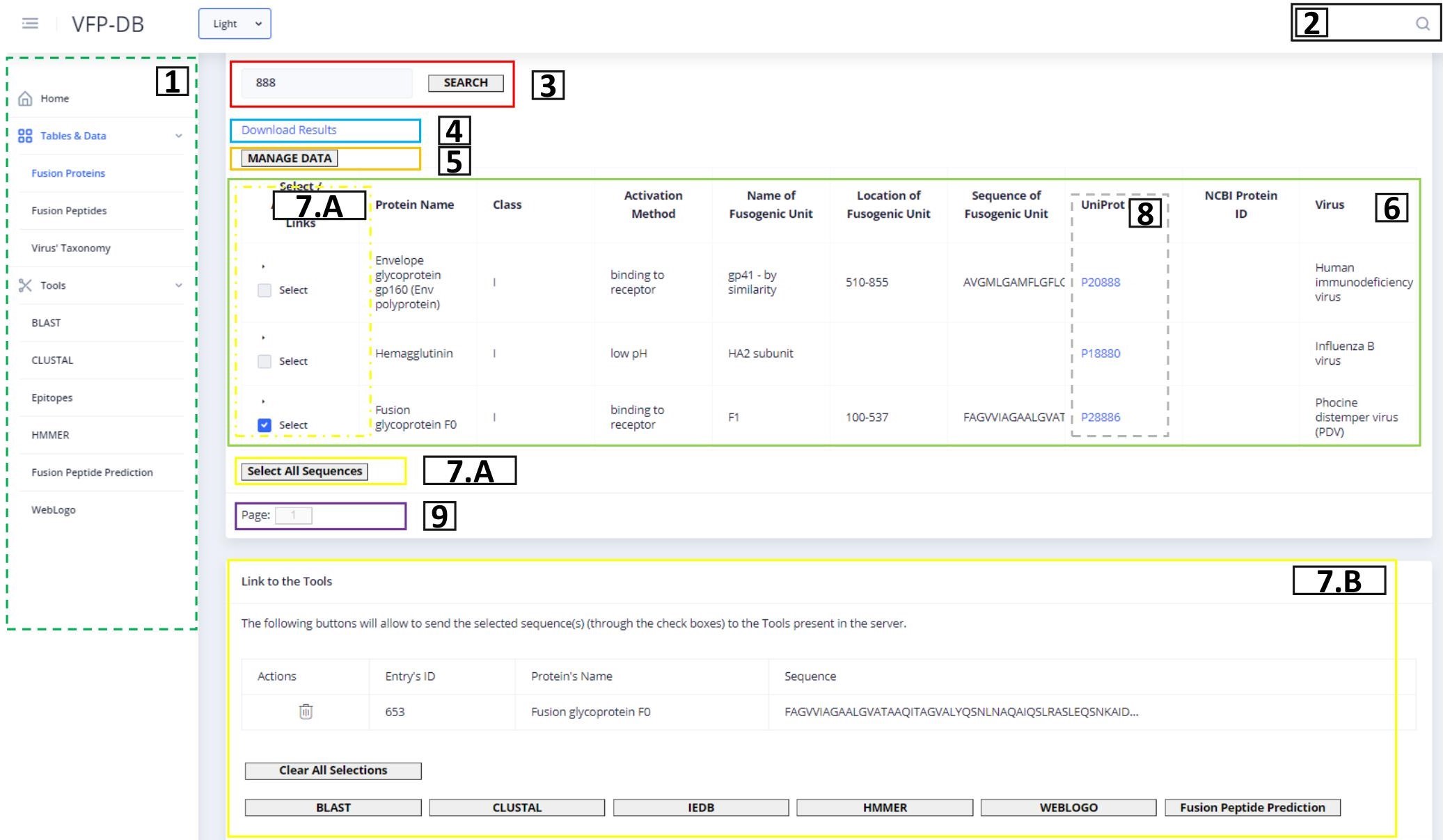}
    \caption{Fusion Protein Page Scheme.} 
    \label{FusionProteinScheme}
\end{figure*}

Regarding ML-based prediction of the location of FPeps within the VFP, this page allows to define a VFP target sequence, as well as specifying which model to use, from the set of available models trained in Propythia: Support Vector Machine, Random Forest, Gradient Boosting, K-Nearest Neighbours, Linear Regression, Gaussian Naïve Bayes and Artificial Neural Network models. Finally, the user can select the window and gap sizes. The page can display those predictions in two types of output: a table which contains every possible peptide probability of being a FPep; and a graphical interface that, for each amino acid, displays a color that represents the maximum probability of all peptides that contain that particular character of being a FPep. Moreover, for each amino acid, a popup box will show up, informing the user about the position within the VFP sequence and the probability value of belonging to a FPep. These results can be downloaded into a text file.

\section{Results}

There are two key components of the front-end that can require an explanation of their elements: the tables and the tools.  

Figure \ref{FusionProteinScheme} shows the representation of the \texttt{Fusion Protein} page. This image contains an example of a table component and will be used to explain all table elements present in our web application. The list above corresponds to the description of the table functionalities from the left to the right and top to down: 

 \begin{figure*}
    \centering
    \includegraphics{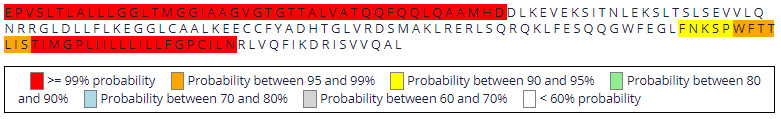}
    \caption{Example of a Machine Learning prediction of the fusion peptide location.} 
    \label{MachinelearningApp}
\end{figure*}

\begin{itemize}
\item On the dashed green rectangle on the left (number 1), there is the side menu of the application, where the user can access the three main data tables, as well as the bioinformatics tools present in the server.
\item On the top right, in the black box (number 2), the user can find the link to the global search tool; upon entering a search term, a pop-up window will be displayed if the main pages contain results.
\item The red rectangle (number 3) contains the search methods previously described, that will determine what is shown in the table; the search box contains an auto-complete system with term suggestions that is executed after the input of 3 characters.
\item The blue rectangle (number 4) allows to download the query result into a CSV file.
\item The orange rectangle (number 5) redirects to the back-end’s administrator page.
\item The green rectangle (number 6) is the table with results from the query request determined by the search term and the current page; within each line, there are links to related data (if available).
\item The yellow rectangles (numbers 7.A and 7.B) show the selection of sequences that will be provided to the web application tools, chosen in the checkboxes on the yellow dashed rectangle, or through the “Select All Sequences" button; upon choosing at least one sequence, the rectangle visible at 7.B will show up, allowing the user to delete some (in the table) or all (in the “Clear All Sequences” option) of the selected sequences and redirect the chosen sequences to the Tools, accessible by the buttons on the bottom of the page.
\item Some columns like the one inside the dashed grey rectangle (number 8) will contain links to redirect to the original source of the data; the links can be UniProt, NCBI Protein, PDB or NCBI Taxonomy data entries.
\item The purple box buttons (number 9) correspond to the pagination methods that will also determine the content of the table.
\end{itemize}

\begin{figure}
    \centering
    \includegraphics[width=\linewidth]{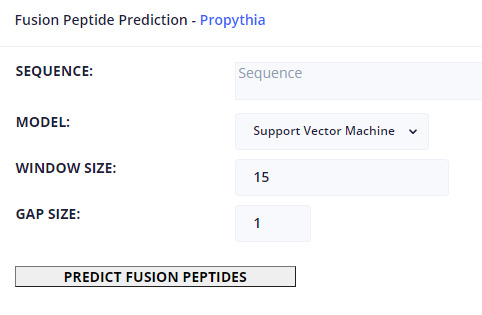}
    \caption{Machine Learning Page.} 
    \label{MLPage}
\end{figure}

Figure \ref{MLPage} shows an example of how all tool pages are structured. Regarding the available tools, they contain a text box to insert the sequence(s) input, which can be auto-filled if protein sequences were selected in the \texttt{Fusion Protein} page: some tools (e.g., the multiple alignment tools) require a minimum of 3 protein sequences in a FASTA format; however, the most basic input of the text box is a single protein sequence. To instruct the user about the type of input that each tool requires, the text boxes contain placeholders with a small description of expected input. After the text box is filled, the user can set custom values for each of the rest of the parameters. The results of the tools can appear in new tabs with text results (e.g using the Clustal or Weblogo), or to access other web portals (e.g., BLAST or HMMER). The results can also appear within the same page, like in the case of ML prediction (as shown in Figure \ref{MachinelearningApp}) and the image result of Weblogo built from the back-end's functions.

\section{Discussion}

This work provides a useful resource for the scientific community working with viruses and their proteins, since it presents a tool that can be used by any researcher interested in obtaining several types of information regarding VFPs.

Currently, the application implements several useful functionalities: it provides an easy access to the MySQL database containing VFP information through a user-friendly web interface. The proposed bioinformatics tools are implemented in the application, through web interfaces. ML models capable of predicting FPeps within the VFP sequence are also available. Finally, a full manual curation of the database information was performed.

Future tasks within the website will include improving the tool by improving current features and adding new ones, including the possibility to define more parameters within the application's tools, improving the initial loading time of the website and the app's aesthetic appearance.

Alongside these tasks, to improve the utility of this application for the scientific community, we will try implement tools capable of predicting the most conserved regions within proteins, which will allow to complement the ML output, due to FPeps tending to be conserved regions. ML models will be improved as new data regarding FPeps is discovered and added to the database, allowing us to update and expand its training dataset. 

Regarding the database's information, it is imperative to add all the relevant new entries published after the submission of this manuscript, hopefully by implementing methodologies capable of performing this type of search automatically and periodically. We are going to add methods that allow users to insert new data, as well as update some inconsistencies that they might find in the application. We aim to add information regarding receptors of VFPs, as well as each VFP’s known cleavage sites. Finally, it is important to redo the automated searches of known inhibitors / antibodies within Pubmed and PDB in a way that can take into consideration the virus' strain.

\section*{Abbreviations}
\textbf{FPep} - Fusion Peptide;
\textbf{ML} - Machine Learning;
\textbf{VFP} - Viral Fusion Protein.

%\begin{acks}
	%Special thanks to Rúben Rodrigues by being the main help in the development of the webserver and the deployment into the server. Another special appreciation to Ana Marta Sequeira, for providing much needed support to understand the Propythia package and to provide machine learning models and datasets.
%s\end{acks}

\section*{Funding}
    This work is funded by COMPETE 2020, Portugal 2020 and FCT - Fundação para a Ciência e a Tecnologia, under the project "Using computational and experimental methods to provide a global characterization of viral fusion peptides", through the funding program "02/SAICT/2017 - Projetos de Investigação Científica e Desenvolvimento Tecnológico (IC\&DT)", with the reference "PTDC/CCI-BIO/28200/2017". This work was also financially supported by Project LISBOA-01-0145-FEDER-007660 (Microbiologia Molecular, Estrutural e Celular) funded by FEDER funds through COMPETE2020 — Programa Operacional Competitividade e Internacionalização (POCI) and by national funds through FCT — Fundação para a Ciência e a Tecnologia.  

\bibliographystyle{ACM-Reference-Format}
\bibliography{sample}

\end{document}